\documentclass[a4paper]{jpconf}
\usepackage{amsfonts}
\usepackage{amsmath}
\usepackage{graphicx}
\usepackage[table]{xcolor}
\usepackage{booktabs}
\usepackage{bm}
\usepackage{bbold}
\usepackage{dsfont}
\usepackage{bbm}

\RequirePackage{mathptm}

\newcommand{\ip}[2]{\langle \,{#1},\,{#2}\,\rangle}

\newcommand{\ro}{\varrho}

\newcommand{\la}{\lambda}
\newcommand{\las}[1]{\lambda_{#1}}

\newcommand{\I}{\mathds{1}}
\newcommand{\conj}[1]{\overline{#1}}

\newcommand{\C}{\mathds C}
\newcommand{\R}{\mathds R}
\newcommand{\M}{\mathds M}
\newcommand{\B}{\mathds B}

\newcommand{\PP}{\mathds P}

\newcommand{\fE}{\mathcal E}
\newcommand{\cA}{{\mathcal A}}
\newcommand{\cB}{{\mathcal B}}
\newcommand{\cM}{\mathcal M}

\newcommand{\cK}{\mathcal K}

\newcommand{\cT}{\mathcal T}

\newcommand{\tra}{\mathrm{tr}\,}
\newcommand{\ptr}[1]{\mathrm{tr}_{#1}}

\begin{document}
\title{Determining quantum correlations in bipartite systems - from qubit to qutrit and beyond }

\author{Andrzej Frydryszak\footnote{Talk given by A.F.}, Lech Jak\'{o}bczyk, Piotr {\L}ugiewicz}

\address{Institute of Theoretical Physics, University of Wroclaw, Poland}

\ead{andrzej.frydryszak@ift.uni.wroc.pl}

\begin{abstract}
We advocate the step change in properties of discrete $d$-level quantum systems, between $d=2$ and
$d\geq 3$. Qubit systems, or multipartite systems containing qubit subsystem, are exceptional in  their
relative simplicity. One faces a step in complexity in valuating measures of quantum correlations for
qutrits and then other higher dimensional qudits. There is a growing number of arguments leading to such conclusion: recently found no-go theorem for generalization of the Peres-Horodecki's PPT criterion \cite{sko}, change in geometry of  state spaces of qubit and higher degree qudits (the so called 'generalized Bloch ball' is not a ball anymore),  restricted possibilities for diagonalization of correlation matrices for bipartite systems, more difficult way for handling the set of relevant families of orthogonal projectors.
\end{abstract}
\section{Introduction}
Quantum correlations in finite dimensional quantum systems are new resources which can fuel quantum
information and computing. Their quantification is crucial, but difficult task. Its complexity depends on
the chosen correlation, definition of its measure and dimension $d$ of the bipartite $d$-level system.
Generic statement valid for all known quantum correlations is: pure states can be uncorrelated or
entangled. Historically first  and mostly studied  correlation is quantum entanglement of qubits,
extended also to qutrits \cite{cm} and then for qudits \cite{rmndmc}, but only for limited types of
states.  Results known for qubit systems can be generalized to some extend to a qubit-qudit case. In
that case a lower dimensional system sets the level of the complexity of the problem. An interesting
aspect of finding the value of correlation measure is a possibility of relating it to the mean values of a
selected observables of  a system. One example of such relation is given for qubits, where entanglement
measure can be expressed in terms of the mean value of spin \cite{fts}.

The correlation we want to focus here on  is the quantum discord.  The general definition of its measure,
while it distinguishes  the quantum and classical character  of correlations in compound systems,
is hardly operational, even for qubit systems. That is why  we restrict our considerations to the so called
measurement-induced one sided quantum geometric discord (MIQGD). It is a version of the geometric
measure of quantum correlations related to the distance used in definition. To have measure
contractive under completely positive trace preserving maps we chose trace norm (Schatten 1-norm).
Such distance used in definition produces proper quantum correlation measure in contrast to the
Hilbert-Schmidt distance. We shall discuss the MIQGD for various $d$-level systems, $d=2,3,\dots $.
However, properties of bipartite systems with $d\geq 3$ change  with restpect to $d=2$. For two-qutrit
states there is no finite set of criteria  of separability   \cite{sko}, what means that there is no
extension of the Peres-Horodecki's necessary and sufficient PPT-criterion valid for qubit-qubit and
qubit-qutrit systems\cite{per, hor1} to higher level systems.

\section{Single qudit system quantum state space}

In the  commonly used notation for $su(d)$ Lie algebras we can describe the one-partite states by the set of generators (we use here uniform notation, where usually for $su(2)$, $\las{j}=\sigma_j$)
\begin{equation}
\tra\, \las{j}=0,\quad \tra\, (\las{j}\las{k})=2\,\delta_{jk},
 \quad \mbox{and} \quad \las{j}\las{k}=\frac{2}{d}\,\delta_{jk}\,\I_{d}
+\sum\limits_{l}\,(d_{jkl}+i\,f_{jkl})\,\las{l}\label{lajlak}
\end{equation}
where $j,k=1,\ldots,d^{2}-1$; the totally symmetric and antisymmetric, respectively, structure constants $d_{jkl}$ and $f_{jkl}$
are given by
$
d_{jkl}=\frac{1}{4}\,\tra\,([\las{j},\las{k}]_{+}\,\las{l})\label{d}
$
and
$
f_{jkl}=\frac{1}{4i}\,\tra\,([\las{j},\las{k}]\,\las{l}).\label{f}
$
\begin{itemize}
  \item $su(2)$:
  $$
  d_{ijk}=0,\quad f_{ijk}=\epsilon_{ijk}
  $$
  \item  $su(3)$:
  $$
  d_{ijk}=
  \left\{
                \begin{array}{ll}
                  \frac{1}{2} &\quad \text{for} \quad (ijk)= (146),\; (157), \; (256), \; (344), \; (355),\\
                  -\frac{1}{2} &\quad \text{for} \quad (ijk)= (247),\; (366), \; (377),\\
                  \frac{1}{\sqrt{3}} &\quad \text{for} \quad (ijk)= (118),\; (228), \; (338),\\
                  - \frac{1}{\sqrt{3}} &\quad \text{for} \quad (ijk)= (888),\\
                  -\frac{1}{2\sqrt{3}} &\quad \text{for} \quad (ijk)= (448),\; (558), \; (668), \; (778)
                \end{array}
              \right.
  $$
  and
  $$
  f_{ijk}=
  \left\{
                \begin{array}{ll}
                  1 &\quad \text{for} \quad (ijk)= (123),\\
                  \frac{1}{2} &\quad \text{for} \quad (ijk)= (147),\; (246), \; (257), \; (345), \; (516), \; (637),\\
                  \frac{\sqrt{3}}{2} &\quad \text{for} \quad (ijk)= (458),\; (678).
                 \end{array}
              \right.
  $$

\end{itemize}

Using these structure constants  one can
introduce the $\star$-product and $\wedge$-products of
  $\R^{d^{2}-1}$ vectors. For $n,\, m\in \R^{d^{2}-1}$ we define
\begin{equation}
(n\star m)_{j}=\sqrt{\frac{d(d-1)}{2}}\,\frac{1}{d-2}\,\sum\limits_{k,l}\,d_{jkl}n_{k}m_{l}
\quad \mbox{and}\quad
(n\wedge
m)_{j}=\sqrt{\frac{d(d-1)}{2}}\,\frac{1}{d-2}\,\sum\limits_{k,l}\,f_{jkl}n_{k}m_{l}
\end{equation}
Let $\la=(\las{1},\ldots,\las{d^{2}-1})$ and
\begin{equation}
\ip{n}{\la}=\sum\limits_{j}n_{j}\las{j}
\end{equation}
The set of states $\ro$ of $d$ - level system i.e. the set of  hermitian of a unit trace matrices which are positive definite is customarily parameterized in the following way
\begin{equation}
\ro=\frac{1}{d}\,\left(\I_{d}+\sqrt{\frac{d(d-1)}{2}}\,\ip{n}{\la}\right),\quad
n\in \R^{d^{2}-1},\label{state}
\end{equation}
where the components of the vector $n$ are
\begin{equation*}
n_{j}=\frac{d}{\sqrt{2d(d-1)}}\,\tr\,(\ro\,\la_{j}),\quad
j=1,\ldots,d^{2}-1.
\end{equation*}
For arbitrary $d$-level system,  pure states get simple characterization
\begin{equation}\label{pure}
\ip{n}{n}=1\quad\text{and}\quad n\star n=n.
\end{equation}
i.e. for solutions of above conditions relevant $\ro$ is a projector.
Only in the case of qubits the $\star$-product is trivial and above condition select the boundary of the
Bloch ball with maximally mixed state laying in its center. Dimension of the quantum  state space  $\mathrm{Q}_d$ is
$d^2-1$ and only for the $d=2$ one gets direct description of the set of states as the ball embeded in
$\mathds{R}^3$. For higher $d$ this set has much richer structure and is highly difficult to characterize
explicitly, what is related also to the fact, that  characterization of the positive-definiteness of  $\rho$
in terms of vector $n$  for arbitrary $d$-level system gets complicated and, in fact, is  not known for
generic case. Nevertheless, for the $d>2$ the frequently adopted name
in literature for this set is "generalized Bloch ball" (GBB), what shouldn't be taken literary. A generic
property of the GBB is that as a convex set it has boundary  placed in between two spheres, the
outsphere of radius $R_d$ and the insphere of the radius $r_d$, where in the Hilbert-Schmidt distance  \cite{bwz}
\begin{equation}
 R_d= \sqrt{(d-1)/2d}, \quad\quad r_d=\frac{1}{\sqrt{2d(d-1)}},
\end{equation}
what means that
$$
R_d=(d-1)r_d.
$$
As stated in \cite{bwz} the $\mathrm{Q}_d$ has a constant height, in the sense that
\begin{equation}\label{height}
  \frac{Ar_d}{V}=d^2-1,
\end{equation}
where $A$ denotes an area of the boundary of the $\mathrm{Q}_d$ and $V$ is its volume. More precisely, it was shown in \cite{sbz} that
such  property holds for the convex set of separable two qubits states and the set of positive partial transpose states of an arbitrary bipartite system.
Let us comment on the generalized Bloch ball $\mathrm{Q}_d$ for  $d=2,3$ .

\subsection*{Bloch ball}
The $\mathrm{Q}_2$ is the simplest  quantum state space and its geometry is the best known. Here $R_d=r_d$ and boundary  of $\mathrm{Q}_2$, $\partial \mathrm{Q}_2$ is identical with the boundary of outer/inner sphere. Pure states are characterized by $<n,n>= 1 $. The condition   $n\star n=n$ in this case is not present.
\subsection*{Single qutrit state space}
Explicit description of the qutrit quantum  state space $\mathrm{Q}_3$ and its visualization  has been intriguing for some time. As for all $\mathrm{Q}_d$ this is a question of describing a convex set with boundary
placed between inner and outer spheres,  here $R_3=2 r_3$. In the Ref. \cite{bwz} one can find what objects cannot serve as a model of $\mathrm{Q}_3$ and then what information one gets using two-dimensional projections and cross-sections. The question of the 'visualization' of the $\mathrm{Q}_3$ was firstly discussed in Ref. \cite{goy-sim} and recently published \cite{gsss}. Tree dimensional 'visualizations' can be obtained by means of the other parametrizations of the $\mathrm{Q}_3$. As shown in Ref. \cite{kklm},  instead of the vector $n$ one can produce a graphical representation of a qutrit  using a  three dimensional vector $a$ and a
metric tensor $\eta$  with distribution of eight independent parameters into 3+5 respectively.
Qutrit states are described by $a$ and $\eta$  such that,   $\eta \cdot a \leq 1$.

Using the description of states by means of the vector $n$, we know, that pure states should satisfy conditions (\ref{pure}), what means that such states are scattered on the outer sphere and discrete rotation is needed to map them geometrically. This is result of the $n \star n = n$ condition and non-triviality of the $d_{ijk}$ constants for the $su(3)$ algebra (for qubits arbitrary rotations are allowed).

\subsection*{Single qudit state space}
Despite some generic properties, there is very little known about generalized Bloch ball for $d\geq 4$.
As described in the Ref. (\cite{bwz}) it is easier to enumerate what characteristics that convex set does
not have. Let us quote here selected properties o the GBB for qudits from the list given in the Ref. (\cite{bwz}):
\begin{itemize}
\item  $\mathrm{Q}_d$ has the 'no hair' property i.e. it is a $d^2-1$ dimensional convex set topologically equivalent to to a ball and it does not have parts of lower dimension.
\item  $B_{r_d} \subset \mathrm{Q}_d \subset B_{R_d}$, where $B_{r} $ denotes a ball with the radius $r$.
\item $\partial \mathrm{Q}_d$ is $d^2-2$ dimensional and contains all states with rank smaller then maximal
\item The set of pure states is  $2d-2$ dimensional and connected. It has zero measure with respect
to $\partial \mathrm{Q}_d$.
\item $\mathrm{Q}_d$ has constant hight, cf. Eq.(\ref{height}).
\end{itemize}
\section{Bipartite qudit system state space}
\subsection*{Two qubits}
The simplest bipartite quantum system in which we can discuss quantum correlations is a system of two qubits.The  total space of states is equal to $\mathrm{Q}_{4}$ which detailed geometrical structure is not known. To get some insight into the geometry of $\mathrm{Q}_{4}$, one can consider
some lower dimensional sections of this set. In \cite{js} two - dimensional sections of the corresponding space of generalized Bloch vectors were considered. On the other hand, the problem of discrimination between separable and entangled states is in this case completely solved. The separability condition based on the notion of partial transposition \cite{per, hor1} is simple and effective. Applied to the above mentioned sections of the set of states, this condition gives some information about geometrical shapes of the sets of separable and entangled states \cite{js}.
\subsection*{Two qutrits and beyond}
In the case of two qutrits, the geometry of the space $\mathrm{Q}_{9}$ is yet more complicated. Some aspects of this geometry was studied in the class
of so called Bell - diagonal states which form a simplex living in the nine - dimensional real linear space
\cite{bhn}. This analysis was then extended to the case of general qudits \cite{bhn1}. What mainly
differs qutrits from qubits is that in the system of two qutrits separability condition based on partial
transposition is not sufficient. It only shows that the states which are not positive after this operation
(NPPT states) are entangled. It turns out that all entangled states can be divided into two classes: free
entangled states that can be distilled using local operations and classical communication (LOCC); bound
entangled states for which no LOCC strategy can be used to extract pure state entanglement
\cite{hor2}.  Last but not least recourse, from the pragmatic point of view, might be the Monte Carlo
sampling of the quantum state space \cite{shang}. It allows obtaining high-quality random samples of
quantum states from higher dimensional $\mathrm{Q}_{d}$, respecting the relevant target distributions
and allowing to evaluate global extremum of a given correlation measure function. This approach is still
to be applied to the geometric quantum discord measures.

\subsection*{New difficulties for higher dimensional systems}
Let us point out some difficulties emerging in systems with higher $d$:
\begin{enumerate}
  \item Considerably more complex structure of the set of separable bipartite states, quantum-classical states etc.
  \item Change in structure of universal enveloping algebra for $su(d)$. Vanishing $d_{ijk}$ symmetric structure constants for $su(2)$ became nontrivial for $d\geq 3$ and modify the geometry of state space via $\star$-product.
  \item Relatively 'shrinked' orthogonal subgroup $R(G)$ originating as the adjoint representation from the unitary transformations $G=\;SU(d)$, $dim\; SU(d)=d^2-1$ compared to  $dim\;O(d^2-1)$ ( {\bf Table 1}).
      \begin{table}[h]
      \caption{\label{dimensions} Dimensions of orthogonal subgroups.}
  \begin{center}
      \begin{tabular}{*4c} \br 
         $G$ & $dim\; G$ & $dim\; R(G)$  & $dim \; O(d^2-1)$\\ \mr 
        SU(2) & 3 & 3 & 3 \\
        SU(3) & 8 & 8 & 28 \\
        SU(4) & 15 & 15 & 105 \\
        \dots & \dots & \dots & \dots \\
        SU(d) & $d^2-1$ & $d^2-1$ & $\frac{1}{2}(d^2-1)(d^2-2)$ \\ \br 
      \end{tabular}
\end{center}
\end{table}
Only for qubits we have the same dimension of these groups and $SU(2)$ is just universal double cover of $SO(3)$.
  \item No possibility to diagonalize all correlation matrices ie. emerging  new sectors in comparison to the qubit intuition.
  \item Steep curve of growing complexity; hopeless perspective to perform effectively  minimization and produce analytical formulas for correlation measures for arbitrary state.
  \item For qubits various correlation measures are equivalent, but split for $d>3$.
 \end{enumerate}

\section{Quantifying quantum correlations in bipartite systems - from qubits to qutrits and beyond }

Consider now two qudits $\cA$ and $\cB$. It is convenient to parametrize the set of states of composite
system as follows
\begin{equation}
\ro=\frac{1}{d^{2}}\left(\I_{d}\otimes
\I_{d}+\sqrt{\frac{d(d-1)}{2}}\,\ip{x}{\la}\otimes \I_{d}+\I_{d}\otimes
\sqrt{\frac{d(d-1)}{2}}\,\ip{y}{\la}
+\sum\limits_{k=1}^{d^{2}-1}\ip{\cK\,e_{k}}{\la}\otimes\ip{e_{k}}{\las{k}}\right)
\label{9state}
\end{equation}
where $x,\, y\in \R^{d^{2}-1}$ and $\{e_{k}\}_{k=1}^{d^{2}-1}$ are the vectors of canonical orthonormal basis of $\R^{d^{2}-1}$. Notice that
\begin{equation*}
x_{j}=\frac{d}{\sqrt{2d(d-1)}}\,\tr\,(\ro\,\las{j}\otimes\I_{d}),\quad
y_{j}=\frac{d}{\sqrt{2d(d-1)}}\,\tr\,(\ro\,\I_{d}\otimes \las{j})
\end{equation*}
and the correlation matrix $\cK$ has elements
\begin{equation*}
\cK_{jk}=\frac{d^{2}}{4}\,\tr\,(\ro\las{j}\otimes\las{k}).
\end{equation*}
The parametrization (\ref{9state}) is chosen is such a way, that the marginals
$\ptr{\cA}\ro$ and $\ptr{\cB}\ro$ are given by the vectors $x$ and $y$ as in (\ref{state}).

\subsection*{Measurement-induced geometric discord}

Let us assume that bipartite system $\cA\cB$ is prepared in a state $\ro$.
Any local measurement on the subsystem $\cA$ will disturb almost all states $\ro$.
This observation yields the definition of a measure of quantum discord.
The one-sided measurement induced geometric discord is defined as the minimal disturbance induced by
any projective measurement $\PP_{\cA}$ on subsystem $\cA$ \cite{dak-ved, pau-oli, jfl}. A distance in the set of states is given by
the trace norm. Namely,
\begin{equation}
{D}_{1}(\ro)=\min\limits_{\PP_{\cA}}\,||\ro-\PP_{\cA}(\ro)||_{1}, \quad
||\sigma||_{1}=\tr\,|\sigma|.
\end{equation}
Local projective measurement $\PP_{\cA}$ is
given by the one-dimensional  projectors $P_{1},\, P_{2},\ldots,\,
P_{d}$ on $\C^{d}$, such that
\begin{equation}\nonumber
P_{1}+P_{2}+\cdots + P_{d}=\I_{d},\quad
P_{j}P_{k}=\delta_{jk}\,P_{k},
\end{equation}
$$
 \PP_{\cA}=\PP\otimes {id},
$$
where
\begin{equation}\nonumber
 \PP(\sigma)=P_{1}\,\sigma\, P_{1}+P_{2}\,\sigma\, P_{2}+\cdots +P_{d}\,\sigma\,P_{d}.
\end{equation}
By $P^0_k$ we shall denote canonical projections on standard orthonormal basis in $\C^d$ and
respective projections on orthogonal complement spaces by $\cM_{0}=\I-P_{0},\quad \cM=\I-P$.
The disturbance of the state after the measurement $\PP_{\cA}$ can be written as
\begin{equation}
S(\cM)\equiv \ro-\PP_{\cA}(\ro)=\frac{1}{d^{2}}\,\big[
\sqrt{\frac{d(d-1)}{2}}\ip{\cM x}{\la}\otimes
\I_{d}+\sum\limits_{k}\ip{\cM\cK\,e_{k}}{\la}\otimes
\ip{e_{k}}{\la}\big] \label{dif}
\end{equation}

\begin{equation}
D_{1}(\ro)=\frac{d}{2(d-1)}\,\min\limits_{\cM}\,\tr\,\sqrt{Q(\cM)}, \quad \quad Q(\cM)\equiv S(\cM)S(\cM)^{\ast},
\end{equation}
where minimum is taken over all $M$ corresponding to measurements on subsystem $\cA$.

\subsection*{Simplifications}
Even such defined measure of quantum discord, more operational then general one, is still difficult to
calculate and some simplifying assumptions are necessary. Let us consider
class of locally maximally mixed states $\ro$
\begin{equation}
\ptr{\cA}\ro=\frac{\I_{d}}{d},\quad \ptr{\cB}\ro=\frac{\I_{d}}{d}.\label{MMM}
\end{equation}
In the chosen parametrization  this corresponds to $x=y=0$ what results in simplified states of the form
\begin{equation}
\ro=\frac{1}{d^{2}}\left(\I_{d}\otimes \I_{d}+\sum\limits_{j=1}^{d^{2}-1}\ip{\cK e_{j}}{\la}\otimes \ip{e_{j}}{\la}\right)
\label{lmm}
\end{equation}
Note that the set of correlation matrices defining above states is convex and is contained in the ball
\begin{equation}
\B_{2}=\left\{A\in \M_{d^{2}-1}(\R)\;:\; ||A||_{2}\leq \frac{d}{2}\sqrt{d^{2}-1}\right\}
\end{equation}
Maximally entangled pure states of this class are defined by correlation matrices lying on
the boundary of the ball $\B_{2}$. However, not every  matrix lying on the border corresponds to some state. Detailed characterization of the set of correlation matrices is not known.

For the states under consideration the disturbance after measurement has the form
\begin{equation}
S(\cM)=\frac{1}{d^{2}}\,\sum\limits_{j=1}^{d^{2}-1}\,\ip{\cM\cK\,e_{j}}{\la}\otimes
\ip{e_{j}}{\la},
\end{equation}
and
\begin{equation}
\begin{split}
Q(\cM)=\frac{1}{d^{4}}\,\bigg[&\,\frac{4}{d^{2}}\,\sum\limits_{j}\,\ip{\cM\cK\,e_{j}}{\cM\cK\,e_{j}}\,
\I_{d}\otimes \I_{d}
+\frac{2}{d\,d^{\prime}}\,\sum\limits_{j}\,\ip{\cM\cK\,e_{j}\star \cM\cK\,e_{j}}{\la}\otimes\I_{d}\\
&+\frac{2}{d\,d^{\prime}}\,\sum\limits_{j,k}\,\ip{\cM\cK\,e_{j}}{\cM\cK\,e_{k}}\I_{d}\otimes \ip{e_{j}\star e_{k}}{\la}
+\frac{1}{d^{\prime 2}}\,\sum\limits_{j,k}\,\ip{\cM\cK\,e_{j}\star \cM\cK\,e_{k}}{\la}\otimes \ip{e_{j}\star e_{k}}{\la}\\
&-\frac{1}{d^{\prime
2}}\sum\limits_{j,k}\,\ip{\cM\cK\,e_{j}\wedge \cM\cK\,e_{k}}{\la}\otimes
\ip{e_{j}\wedge e_{k}}{\la}\,\bigg]
\end{split}\label{QM}
\end{equation}
The spectrum of $Q(\cM)$ for arbitrary correlation matrix is rather difficult to obtain, but interestingly
enough, one can find a universal lower bound for $D_{1}$ and for analogous quantum discord measure
based on the Hilbert-Schmidt distance $D_{2}$ \cite{lfj}.
Let $\cK$  defines  the corresponding locally maximally mixed state $\ro$, then
\begin{equation*}
D_{2}(\ro)\geq
\frac{4}{d^{3}(d-1)}\;\Xi(\cK)\quad\text{and}\quad
D_{1}(\ro)\geq \frac{1}{d(d-1)}\;\sqrt{\Xi(\cK)},
\quad\text{where}\quad \Xi(\cK)=\sum\limits_{j=d}^{d^{2}-1}\eta_{j}^{\downarrow}.
\end{equation*}
By the  $\{\eta_{j}^{\downarrow}\}$ we denote the collection of eigenvalues of $\cK\cK^{T} $
taken in non- increasing  order.
Let us observe that states related to correlation matrices with $rank \,\cK\geq d$ have non-zero
quantum discord.

However, to get more specific information one  has to admit further restrictions on the form of the correlation matrices. For  $\cK=t\,V_{0}$, and $V_{0}\in {O(d^{2}-1)}$ one can proof  that \cite{lfj}
\begin{equation}
Q(\cM)=\frac{t^{2}}{d^{4}}\bigg[\frac{4(d-1)}{d}\,\I_{d}\otimes
\I_{d}+\frac{2}{d}\,\I_{d}\otimes \sum\limits_{k}X_{k}\,\las{k}
+\sum\limits_{j,k}Y_{jk}\,\las{j}\otimes \las{k}\bigg]
\label{QMI}
\end{equation}
where
\begin{equation}
X_{k}=\tr\,(\cM\,V_{0}\Delta_{k}V_{0}^{T}),\quad
Y_{jk}=\tr\,(V_{0}^{T}\cM\Delta_{j}\cM V_{0}\Delta_{k}+V_{0}^{T}\cM F_{j}\cM V_{0}F_{k})
\end{equation}
and $(\Delta_{k})_{jl}=d_{jkl},\; (F_{k})_{jl}=f_{jkl}$.
Let us observe that terms involving $Y_{jk}$ are related to $\star$ and $\wedge$ - products in ${\R}^{(d^2-1)}$.

Above form of  the $Q(\cM)$ can be starting point to obtain analytical expression
for trace - norm quantum discord. The additional conditions on the
 correlation matrices one has to impose come from the case study of  the system  of two qutrits
 \cite{jfl, lfj}. Final answer valid for two-qudit systems can be formulated as follows: there two
 families of correlation matrices defining states for which analytical formula for MIQGD can be obtained.
 They are given by correlation matrices of the form
 \begin{equation}
\cK^a=t\,V,\quad V\in R(SU(d))
\end{equation}
and
\begin{equation}
\cK^{aa}=t\,\cT,\quad   \cT= V_{1}I_{0}V_{2}^{T},\; V_{1},\,V_{2}\in R(SU(d)),
\end{equation}
where
\begin{equation*}
\left(I_{0}\right)_{kk}=\frac{1}{2}\tr (\las{k}^{T}\las{k}),\quad k=1,\ldots,d^{2}-1
\end{equation*}
For the first family we obtain
\begin{equation}
Q^{\,{a}}(\cM)=\frac{t^{2}}{d^{4}}\left[\left(\frac{2}{d}\right)^{2}d(d-1)\,\I_{d}\otimes
\I_{d}-2\sum\limits_{k=2}^{d}U\las{k^{2}-1}U^{\ast}\otimes
\tau_{V}(U)\tau_{\cT}(\las{k^{2}-1})\tau_{V}(U^{\ast})\right]
\end{equation}
and for the second one
\begin{equation}
\begin{split}
Q^{\,{aa}}(\cM)=\frac{t^{2}}{d^{4}}\bigg[&\bigg(\frac{2}{d}\bigg)^{2}d(d-1)\,\I_{d}\otimes
\I_{d}+2\,\bigg((d-2)\sum\limits_{k}\las{k}\otimes\tau_{\cT}(\las{k})\\
&+\sum\limits_{k=2}^{d}U\las{k^{2}-1}U^{\ast}\otimes
\tau_{\cT}(U^{\ast})\tau_{\cT}(\las{k^{2}-1})\tau_{\cT}(U)\bigg)\bigg],
\end{split}
\end{equation}
where  $\tau_V$ and $\tau_\cT$ are Jordan automorphisms and antiautomorphism of the $\M_{d}(\C)$ (for details cf. Ref. \cite{lfj}).

For above  classes of states one gets explicit formulas:
\begin{itemize}
  \item[(i)]  for  $\ro\in \fE^{{a}}$
\begin{equation}
D_{1}(\ro)=|t|,\quad -\frac{d}{2(d-1)}\leq t\leq \frac{d}{2(d+1)}
\end{equation}
  \item[(ii)]  for $\ro\in \fE^{{aa}}$
\begin{equation}
D_{1}(\ro)=\frac{2}{d}\,|t|,\quad -\frac{d}{2(d^{2}-1)}\leq t\leq
\frac{d}{2}
\end{equation}
\end{itemize}
It is interesting to compare above families to the two known distinguished classes of states: Werner
states and isotropic states. Condition defining Werner states is
\begin{equation}
\ro = U\otimes U\,\ro\,U^{\ast}\otimes U^{\ast},\quad U\in {SU(d)}
\end{equation}
In turns out that these states fall into the class $\fE^{{a}}$ and property (\ref{wer}) means that
\begin{equation*}\label{wer}
Q^{{a}}(\cM)=U\otimes U\, Q^{{a}}(\cM_{0})\,U^{\ast}\otimes U^{\ast}
\end{equation*}
Condition defining the so called isotropic states is
\begin{equation}
\ro= U\otimes \conj{U}\,\ro\, U^{\ast}\otimes U^{T},\quad U\in {SU(d)}
\end{equation}
These states fall into the class $\fE^{{aa}}$ and  for them
\begin{equation*}
Q^{{aa}}(\cM)=U\otimes\conj{U}\,Q^{{aa}}(\cM_{0})\,U^{\ast}\otimes U^{T}.
\end{equation*}
Hence, in both cases a minimization procedure is not needed.
\section*{Conclusions}
In our work we have presented  some aspects of the  complex problem of finding values of quantum
correlation measures. As an illustration we have discussed the measurement-induced  one sided 
quantum geometric discord based on the trace distance. While, on the one hand we enlist and comment
types of difficulties arising for the higher $d$-level systems and stress the step change between $d=2$
and $d\geq 3$ systems, on the other hand we show that there are important instances, where one
can effectively  avoid troublesome minimization procedure and obtain strict results for the MIQGD.
\section*{References}

\end{document}